\documentclass[11pt]{article}
\pdfoutput=1
\usepackage[T1]{fontenc}
\usepackage[utf8]{inputenc}
\usepackage{amsmath,amssymb}
\usepackage{graphicx}
\usepackage[margin=1in]{geometry}
\usepackage{authblk}
\usepackage[numbers,sort&compress]{natbib}
\usepackage{hyperref}

\title{Neural Networks for Inverse Design of Cascaded-Mode Near-Field Landscapes}
\author[1,3,*]{Wannes Luts De Martelaere}
\author[1]{Joeri Lenaerts}
\author[1,2,*]{Vincent Ginis}
\affil[1]{Data Analytics Lab and Applied Physics, Vrije Universiteit Brussel, Pleinlaan 2, 1050 Brussel, Belgium}
\affil[2]{Harvard John A. Paulson School of Engineering and Applied Sciences, Harvard University, Cambridge, MA 02138, USA}
\affil[3]{Department of Physics, Korea University, Seoul 02841, Republic of Korea}
\affil[*]{Corresponding authors: \texttt{wannes.demartelaere@gmail.com}, \texttt{vincent.ginis@vub.be}}
\date{}

\begin{document}
\maketitle
\begin{abstract} 
Structuring optical near-fields is important for applications in microscopy and nanoparticle manipulation. Traditionally, near-fields are structured using antenna nanostructures that locally convert propagating far-fields into bound near-fields.
Recently, a remote structuring approach was proposed using cascaded mode interference in a multimode waveguide. However, determining the complex coefficients of the optimal modal combination needed to obtain specific near-fields remains a challenge. 
We address this inverse design problem using artificial neural networks. We model the relationship between the design parameters and near-field landscapes using multilayer neural networks. After training, these networks are used for gradient-based optimization to reconstruct target near-field profiles. 
We implement this methodology to design longitudinal and lateral field variations. Our approach designs simple and complex longitudinal landscapes, demonstrating accurate prediction and flexibility. Lateral field reconstruction is more challenging but improved with training data selection and augmentation. 
This work establishes deep learning as an efficient and scalable framework for cascaded-mode near-field inverse design.
\end{abstract}
\section{Introduction}
\begin{figure}[htbp]
\centering
\includegraphics[width=\textwidth
]{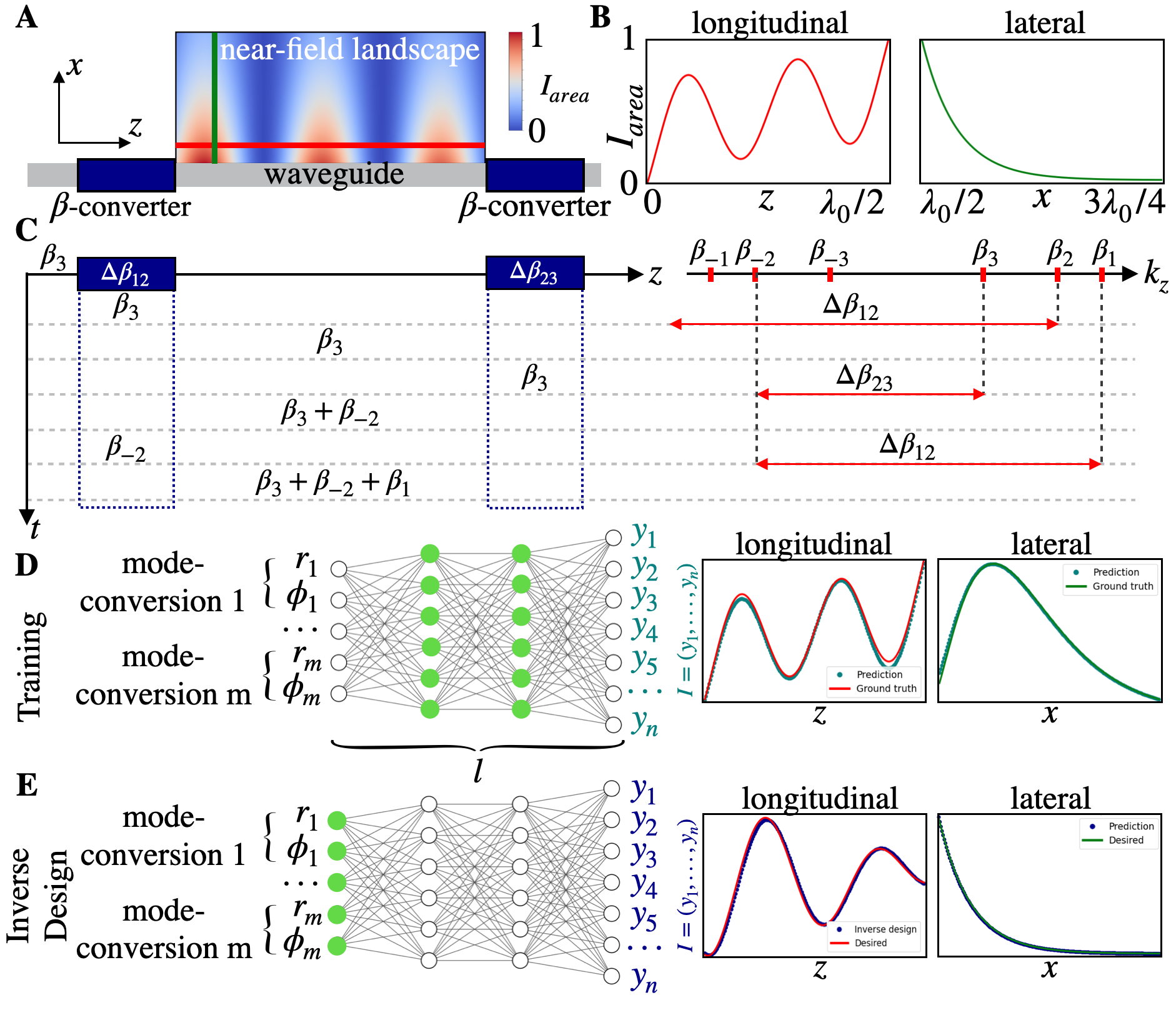} \\
\caption{\textbf{Remote near-field structuring using mode-converters.}
\textbf{(A)} A schematic of a near-field landscaping device consisting of a multimode waveguide and two sets of $\beta$-converters delineating an area of interest. A single mode is fed into the device, and by iterative interactions with the $\beta$-converters a cascade of counterpropagating modes, with different complex amplitudes, is generated. Interference between these counterpropagating modes will engender a specific near-field landscape over the area of interest. By independently controlling the reflectivity ($r_i$) and phase shift ($\phi_i$) induced by each mode conversion, the near-field landscape can be engineered. 
\textbf{(B)} A longitudinal (red) and lateral (green) cross-section of a near-field landscape.
\textbf{(C)} A step-by-step (top to bottom) mechanism showing the formation of a three-mode cascade in both physical (left) and reciprocal space (right). In reciprocal space, $\beta$-converters are represented by vectors bridging counterpropagating modes with different mode indices. Each combination of modes in the cascade results in a specific near-field landscape. Thus when a certain landscape is desired, a method to single out an optimal combination of $\beta$-converters is needed. This can be accomplished using deep learning-based inverse design. 
\textbf{(D)} In the first step a neural network is trained to map a set of reflectivities and phase shifts, defining the combination of $\beta$-converters, to the resultant near-field landscapes (lateral or longitudinal). 
\textbf{(E)} In the second step, the weights and biases of the trained network are fixed. One can now solve for a local optimal set of design parameters using gradient descent to iteratively update the input parameters to better approximate the desired landscape.
}
\label{fig_1}
\end{figure}

Guided modes are solutions to Maxwell's equations for fields exclusively propagating within a confining structure, known as a waveguide. In the region of space surrounding such a guide a non-negligible field---the near-field of the confined modes~\cite{girard2000physics}---will typically exist. In recent years this near-field has been of particular interest for its application in subwavelength microscopy~\cite{novotny2007history,chen2022rapid}, near-field sensing~\cite{novotny2012principles,ginis2019using}, the development of near-field scanning optical microscopes~\cite{ash1972super}, nanoparticle manipulation~\cite{nieto2004near,bekshaev2013mie,bliokh2015spin,liu2018three,bradac2018nanoscale}, dispersion engineering in nanostructured devices~\cite{he2019analogue}, and nonradiative energy transfer~\cite{andrew2000forster,yu2022near}.
Some of these applications require a specific near-field intensity distribution (near-field landscape), necessitating a method for structuring these fields~\cite{ginis2020remote}. To the present day, several techniques have been presented~\cite{fernandez2017enhancing, wei2018toward, piccardo2022roadmap, hu2018experimental}, most of them relying on classical antenna theory where propagating waves are locally converted into specific near-field landscapes using dielectric or plasmonic nano-scatterers~\cite{pohl2012near, kuznetsov2016optically, hsiao2017fundamentals}. The main drawback of these methods is the local conversion, resulting in bulky optics for on-chip applications. Recently, we proposed a new method for the remote rather than local structuring of the near-field~\cite{ginis2020remote}. This method relies on the cascaded mode conversion and interference of counterpropagating modes with different complex amplitudes to structure the near-field at an interface~\cite{ginis2020remote,ginis2023resonators}. 
A schematic representation of the setup used for this remote sculpting method is shown in (Fig. \ref{fig_1}A). Here, an area of interest is delineated by two sets of mode-converters ($\beta$-converters). These $\beta$-converters are used to engineer a specific cascade of counterpropagating modes, by selectively converting and reflecting guided modes back and forth. Interference between the modes results in a specific near-field landscape. Both a longitudinal (along the interface) and a lateral (perpendicular to the interface) cross-section of a near-field landscape are shown in (Fig. \ref{fig_1}B).
Note that $\beta$ is used to refer to the longitudinal component of the guided-mode wavenumber, so $\beta = k_0n_\mathrm{eff}$, with $k_0$ the free-space wavenumber and $n_\mathrm{eff}$ the effective mode index. These longitudinal wavenumbers have a specific spacing in reciprocal space due to the discrete nature of the modes~\cite{mansuripur2002classical, bogaerts2005nanophotonic, bharadwaj2009optical, chuang2012physics, ginis2023resonators}. As a consequence, it is possible to design $\beta$-converters that can reflect one specific mode into another counterpropagating mode, while leaving all other modes virtually unaffected~\cite{ginis2020remote, haas2021design, quaranta2018recent, perez2018mode}. This can be understood best in reciprocal space, where a mode-converter can be pictured bridging the difference between two counterpropagating modes. Thus, the physical properties of the $\beta$-converters entirely determine the selectivity, reflection coefficient ($r$), and phase shift ($\phi$) of each conversion and consequentially the longitudinal wavenumber and complex amplitude of the reflected mode.
A step-by-step mechanism showing the creation of a three-mode cascade in both physical (left) and reciprocal (right) space is shown in (Fig. \ref{fig_1}C). Here the combination of $\beta$-converters is chosen such that when mode $\beta_3$ is incident along the positive $z$-axis, it will interact with the rightmost mode-converter, converting it into the counterpropagating mode $\beta_{-2}=-\beta_{2}$. This mode will interact with the leftmost converter, where it is projected onto mode $\beta_1$, resulting in a three-mode cascade composed of modes with different complex amplitudes. Interference between these modes will result in an electric field of the form 
\begin{equation}
    E_{area} = E_0\left(e^{i\beta_3z}+r_1 e^{i(\beta_{-2}z+\phi_1)}+r_1r_2e^{i(\beta_1z+\phi_1+\phi_2)}\right),
    \label{form:0}
\end{equation}
with $E_0$ the amplitude of the initial incident mode, $r_1$ and $r_2$ the reflection coefficients and $\phi_1$ and $\phi_2$ the phase shifts induced by the respective mode-converters. Note that one complexity here is that the amplitude and phase of each mode are determined by all prior mode conversions.
This process can be expanded for systems with an arbitrary number of modes ($n$). The eventual intensity distribution of the near-field landscape resulting from such a cascade is described by~\cite{ginis2020remote}
\begin{equation}
    \frac{I_{area}}{I_0} = \sum_{i=1}^n\left(\prod_{j=0}^{i-1}r_j^2\right)+2\sum_{l=2}^n\sum_{m=1}^{l-1}\left(\prod_{p=0}^{l-1}r_p\prod_{q=0}^{m-1}r_q \cos\left[(\beta_l-\beta_m)z-\sum_{r=m}^l\phi_r\right]\right),
    \label{form:1}
\end{equation}
with $I_0$ the intensity of the initial incident mode and $\beta_i$ the longitudinal wavenumber, in the order in which they are generated. One complexity so far unuttered is that the reflection coefficients $r_i$ are determined by both the conversion efficiency and the difference between the near-field amplitudes of the interconverted modes at the interface~\cite{ginis2020remote}. 
The first term in Eq.~\ref{form:1} is a constant (DC) contribution. The second term holds the spatial variation and resembles a Fourier series with contributing frequencies as a difference of frequencies, rather than an equidistant set. Nevertheless, it can be shown that many longitudinal spatial profiles can be arbitrarily well approximated by this sum of cosines~\cite{ginis2020remote}. From Eq.~\ref{form:1} it also follows that it is possible to design lateral field distributions that go beyond the classical evanescent decay since modes with different propagation constants $\beta_i$ also have different decay constants $\gamma_i = (\beta_i^2-k_0^2)^{1/2}$. 
So to rehash, Eq.~\ref{form:1} allows one to calculate the optical response for a specific cascade of modes and ergo for a set of design parameters describing the $\beta$-converters used to create the cascade. One can now also ask the inverse question: ``What combination of modes and thus mode-converters is needed to sculpt a desired landscape?''
Multiple computational methods can be used to tackle such an inverse design problem~\cite{molesky2018inverse} like evolutionary algorithms~\cite{wiecha2019design,mayer2022genetic,ginis2020remote}, density topology optimization~\cite{molesky2018inverse} or the adjoint method ~\cite{giles2000introduction, hughes2018adjoint} (mathematically inverting the equation). The evolutionary and density topology algorithms are computationally expensive~\cite{peurifoy2018nanophotonic, lenaerts2020artificial}, requiring exponentially more computation time with increased complexity and number of design parameters. The adjoint method does offer a more efficient method but requires a deep knowledge of the physics and becomes nontrivial for increasingly complex problems~\cite{peurifoy2018nanophotonic}.  
In the field of photonics the use of artificial neural networks for the design of complex photonics components has recently gained traction~\cite{blanchard2020teaching, liu2018generative, kudyshev2020machine, jiang2021deep}. 
Deep learning-based inverse design~\cite{so2020deep, wiecha2021deep, peurifoy2018nanophotonic, liu2021tackling, yao2023deep, ma2021deep} has since emerged as an alternative to solve inverse design problems. This inverse design technique has for example been used for the design of integrated photonic power splitters~\cite{tahersima2019deep}, metasurfaces~\cite{yuan2021efficient}, Fabry-Perot resonators \cite{lenaerts2020artificial} and for performance optimization and inverse design of plasmonic waveguide cavity couplers~\cite{zhang2019efficient} and novel graphene metamaterials~\cite{zhang2020machine}. 
This alternative inverse design approach was introduced into the field of nanophotonics by Peurifoy et al.~\cite{peurifoy2018nanophotonic} for the inverse design of multilayer nano-particles. 
The technique is a two-step method where in a first step, a neural network is trained to map a set of design parameters to the corresponding optical response, in our case the near-field landscape (Fig. \ref{fig_1}D). In a next step, the trained network is used to perform gradient descent on the design parameters. That is to say, the weights and biases of the trained network are fixed, and by means of gradient descent the input parameters are updated such that the network's prediction approximates the desired outcome better and better over consecutive steps, eventually yielding a local optimal set of design parameters, as visualized in (Fig. \ref{fig_1}E). In our work these design parameters describe a cascade of modes or, equivalently, the set of mode-converters.
This deep learning-based inverse design technique offers three main advantages over the other inverse design methods. Firstly it relies on neural networks making the technique inherently scalable, and ideal for solving complex problems. Secondly, once a neural network is trained the gradient can be found analytically instead of numerically, making this inverse design technique significantly faster than the alternatives~\cite{peurifoy2018nanophotonic}. Thirdly, since we use neural networks to map the input and output, we do not rely on an analytical formula~\cite{lenaerts2020artificial, peurifoy2018nanophotonic}, meaning that this method can also be used for problems where no analytical formula is known, if numerical or experimental data is available~\cite{raissi2019physics, chen2020physics}. This technique is thus ideal for a myriad of inverse design problems within photonics. 
In this work we thus opted to employ deep learning-based inverse design as it offers a scalable platform for the design of increasingly complex near-field landscapes, making it ideal for cascades with an increasing number of modes. Moreover as we showed in previous work~\cite{ginis2020remote}, structuring lateral near-field profiles using generic inverse design algorithms does not work well, as a result of the increased complexity resultant from the relative decay constants of the various modes, hence requiring a  new approach.  
\section{Predicting the Longitudinal Near-Field Landscape from the Modes in the Cascade}
\begin{figure}[htbp]
\centering
\includegraphics[width=\textwidth]{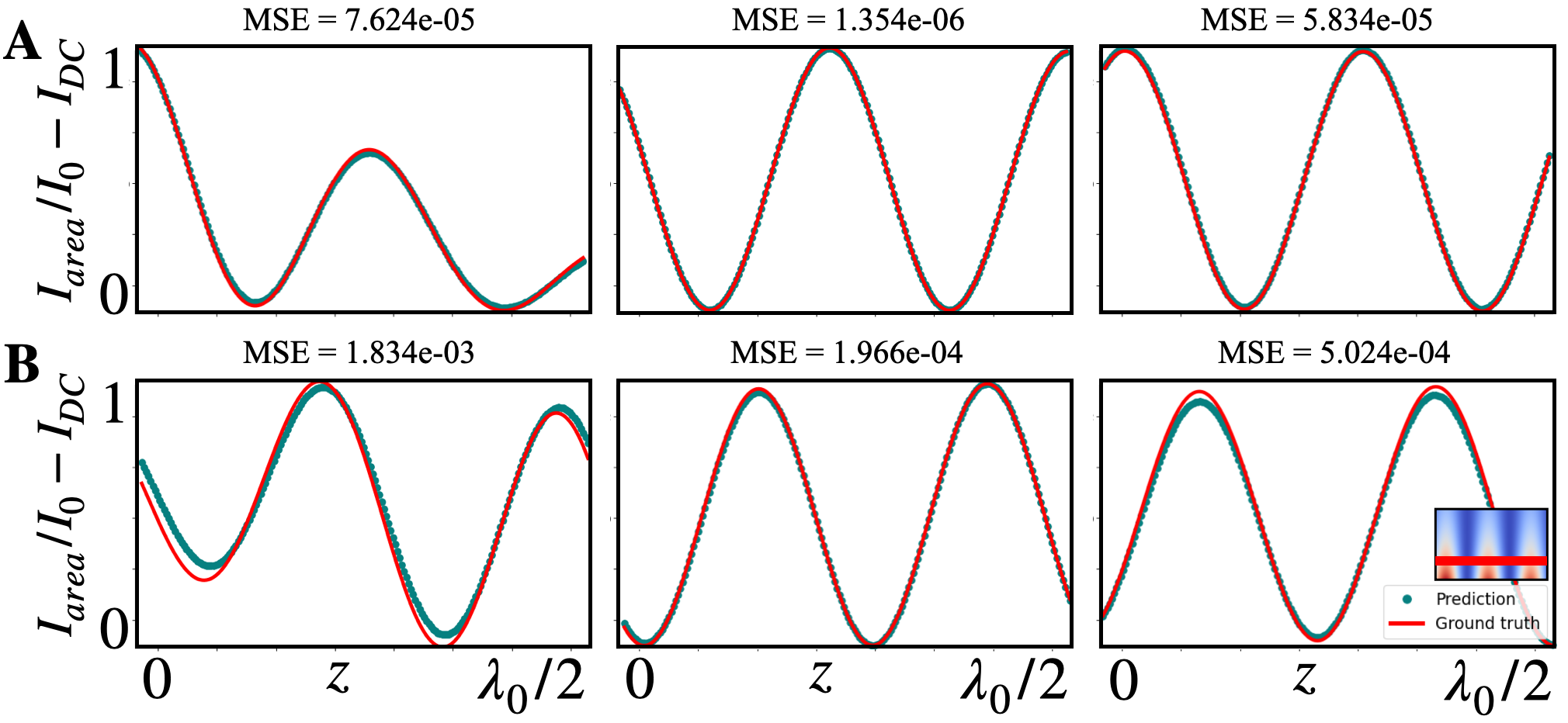} \\
\caption{\textbf{Longitudinally trained Neural Networks}
\textbf{(A)} The performance of a network trained on a four-mode cascade (three conversions). The analytically computed ground truth is shown in red, and the prediction made by the network is shown in green.
\textbf{(B)} The same for the case of a ten-mode cascade (nine conversions).
}
\label{fig_2}
\end{figure}

In order to implement the deep learning-based inverse design technique, a neural network needs to be trained to map the design parameters (which define the combination of $\beta$-converters) to the resultant landscape. To do this, one needs to first decide on the degrees of freedom. That is to say, pick the design parameters that can vary so as to have a sufficiently complex yet computationally manageable system. The available design parameters are the geometry and optical properties of the waveguide, which determine the number of modes and the corresponding wavenumbers, the location and geometry of the area of interest, the initial incident mode ($\beta$, amplitude and phase), the number and order of the different mode conversions and the selectivity, the reflectivity and the phase shift of each conversion. 
In order to reduce complexity, one specific waveguide was chosen thereby fixing the properties of the modes~\cite{chuang2012physics}. Secondly, only a cross-section of the near-field landscape, either lateral or longitudinal was considered. Also, the area of interest was each time predetermined as ($z$ : $0$ to $\lambda_0/2$, $x=\lambda_0/2$) for the longitudinal and ($z=0$, $x$ : $\lambda_0/2$ to $3\lambda_0/4$) for the lateral cross-sections, where $x$ = $\lambda_0/2$ is the surface of the waveguide. The initial incident mode was also fixed and normalized, and no initial phase shift was assumed ($r_0=1$, $\phi_0=0$). Lastly, the number of mode-converters, their selectivity and their placement were fixed for each network. Thus the remaining degrees of freedom were the reflection coefficients and phase shifts for each conversion. By choice of these degrees of freedom, the input of the network was of the form: [$r_1$,$r_2$,...,$r_{n-1}$,$\phi_1$,$\phi_2$,...,$\phi_{n-1}$], for a cascade with $n-1$ mode conversions ($n$ modes). Each of these parameters was normalized to a range [$0$,$1$].
In total, 100,000 of these input arrays were randomly generated. For each, the corresponding output considered was $200$ equidistant points of the spatial variation of the normalized longitudinal intensity profile, solved for using Eq.~\ref{form:1}. The combination of input and output data constituted the mother data pool, which was split into $75\%$ training set, $15\%$ validation set, and $10\%$ test set samples. 
With the training data covered, the actual neural network was constructed. The structure used had $2n-2$ input nodes (corresponding to the input parameters of $n-1$ mode conversions), followed by $6$ layers with $100$ neurons each, and finalized with a $200$-neuron output layer, corresponding to the $200$ intensity points of the corresponding profiles. The Swish activation function~\cite{ramachandran2017searching}, Adam optimizer~\cite{kingma2014adam} ($\eta=0.001$) and mean squared error loss function (MSE) were used during training. The structure of this neural network was modeled after the networks successfully used for the inverse design of nanophotonic components in~\cite{lenaerts2020artificial, peurifoy2018nanophotonic, tahersima2019deep, liu2021tackling}.
Fig. \ref{fig_2} shows the performance of these forward-trained neural networks for predicting longitudinal field profiles given the corresponding design parameters. Fig. \ref{fig_2}A displays the case of a four-mode cascade, for a network trained for $500$ epochs. The ground truth is shown in red and is given by the analytically calculated field (Eq.~\ref{form:1}). The prediction made by the trained network is shown in green. The MSE (between prediction and ground truth) for each sample is displayed above each image, as a way of quantifying the performance of the network. The same was also done for a ten-mode cascade (Fig. \ref{fig_2}B), in which case the training dataset was quadrupled and the training time was doubled in order to improve performance given the increased complexity of the setup.
The performance of these forward-trained networks is important given that the inverse design step is run on these. Any discrepancy induced by a poorly performing neural network will thus translate into a reduced performance of the inverse design step itself. In other words, the performance of the forward-trained neural networks provides an intrinsic limit to the performance of the inverse design step.  
\section{The Inverse Design of Longitudinal Near-Field Landscapes}
\begin{figure}[htbp]
\centering
\includegraphics[width=\textwidth]{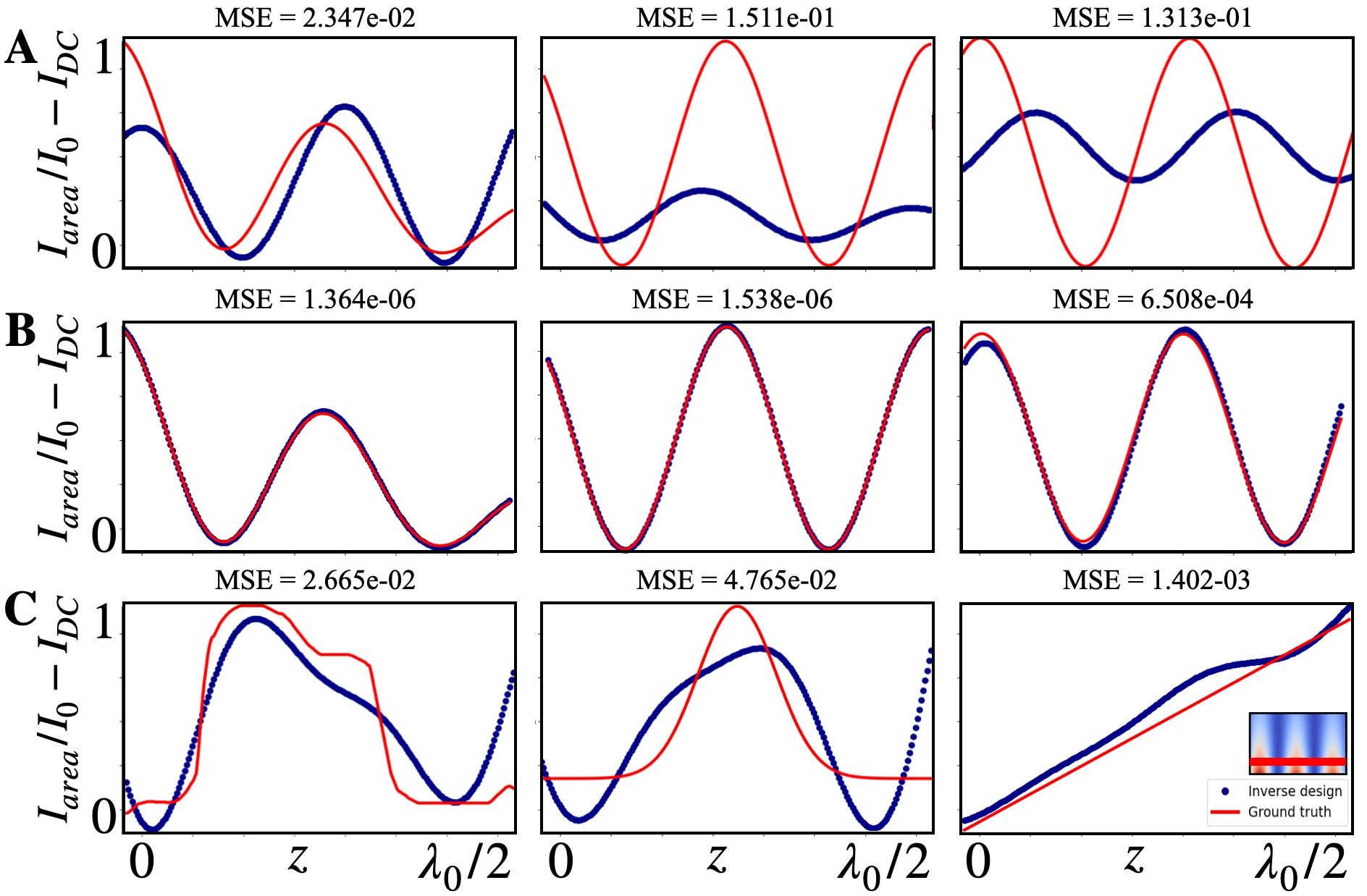} \\
\caption{\textbf{Inverse design of longitudinal fields}
\textbf{(A)} The performance of the inverse design step for landscapes from the test dataset, on a four-mode network when using a random guess for the design parameters and no limitations on the parameter space. The desired field is shown in red and the field resultant from the predicted design parameters is shown in blue. 
\textbf{(B)} The performance can be significantly improved when implementing two modifications. Firstly, one can limit the parameter space to only physically attainable values ($r_i$ in [$0$,$1$]). Secondly, one can also use an inversely-trained network (input: the optical response, output: the design parameters) to make an initial guess. 
\textbf{(C)} Some more complex landscapes, not from the test dataset, were attempted, using the modifications discussed in (B), on a ten-mode network.}
\label{fig_3}
\end{figure}

The deep learning-based inverse design technique (Fig. \ref{fig_1}E) was implemented on the trained neural networks of the previous section (Fig. \ref{fig_1}D). To do this, the weights and biases of the network were fixed and a target near-field landscape, not included in the training data, was then given to the network. This target landscape was taken either from the test dataset or it was artificially generated. The parameters of the input layer, i.e., the design parameters, were initialized to random values. Gradient descent was then applied to the input nodes, using the MSE loss function to compare the network's predicted field to the desired target field. The input parameters were iteratively updated to minimize the MSE loss, thereby reconstructing the target field. This process was repeated over multiple epochs to converge on a local optimal set of design parameters describing the required modal combination.
Fig. \ref{fig_3}A shows some results obtained after running the inverse design step on a four-mode network for $5000$ epochs. These samples were taken from the test dataset; it is thus known that these fields can be shaped using the design parameters considered. Hence the performance here is limited only by the neural network's performance and the inverse design step itself. 
However, as Fig. \ref{fig_3}A shows, this naive inverse design step fails to recover a satisfactory set of design parameters, with the reconstructed field (blue) deviating strongly from the desired target (red). Since the neural network itself was observed to perform well (Fig. \ref{fig_2}), this poor reconstruction must originate in the inverse design step itself.
Two modifications were made to improve these results. Firstly, the parameter space the inverse design step could operate in was limited to only the physically attainable space. This was done by artificially setting a parameter equal to the nearest allowed value when a nonphysical result was returned, like a negative value for the reflection coefficients or a larger than one reflection coefficient. This significantly improved the performance. 
A second modification made was to use an inversely-trained neural network (trained from landscape to design parameters) to make the initial guess. In general, and also observed here, such a neural network does not perform very well since it has to learn a non-one-to-one mapping where the same optical response can be generated by multiple combinations of design parameters. This non-uniqueness problem is hard for a neural network to solve \cite{lenaerts2020artificial, liu2018training}, hence the need for the inverse design step on the forward-trained network. However, we did find that using the output of this inversely-trained network did in many instances make for a good initial guess for the design of longitudinal landscapes.  
With these modifications in place, the inverse design step was run again on the same fields as shown in (Fig. \ref{fig_3}A), yielding much-improved results (Fig. \ref{fig_3}B). To assess the limits of this inverse design based near-field sculpting method, we also tried a ten-mode cascade for the design of some more complex artificial field distributions like an Elephant-curve, a Gaussian and a Bisector (Fig. \ref{fig_3}C). Note that these fields do not come from the test dataset.
Therefore, it is unknown whether they can be effectively shaped, given that the design freedom is constrained by the considered cascade of modes.
The performance here thus is not just a function of the inverse design step itself; it is also intrinsically limited by the design freedom provided by the cascade of modes. With these figures, we intend to convey the performance not of the inverse design step alone, but of the combined framework: the inverse design of near-field landscapes using the cascaded-mode interference sculpting method.
\section{Predicting the Lateral Near-Field Landscape from the Modes in the Cascade}
\begin{figure}[htbp]
\centering
\includegraphics[width=\textwidth]{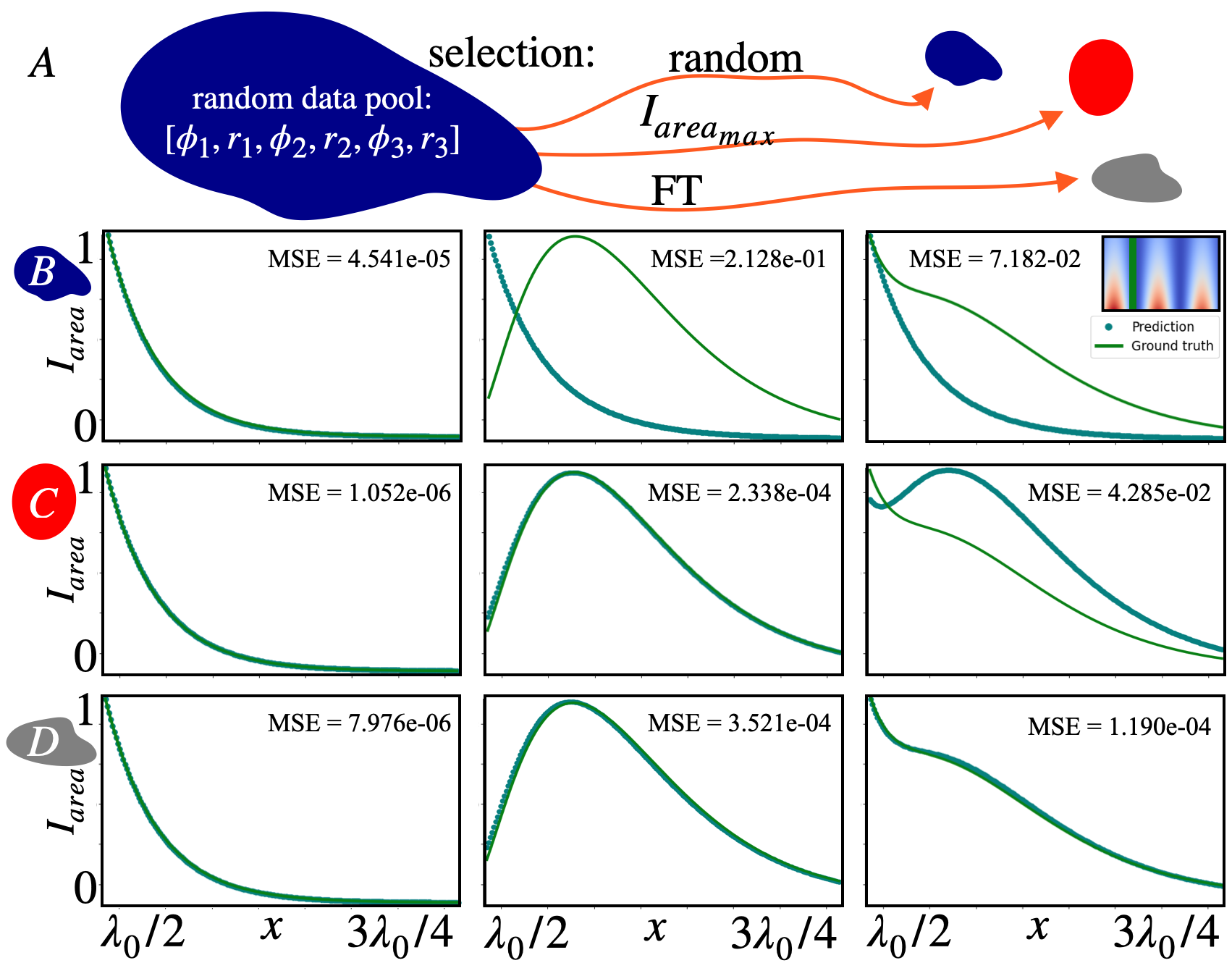} \\
\caption{\textbf{Laterally trained networks} 
\textbf{(A)} To successfully train neural networks for the lateral field design, one needs to bias the training data to sufficiently expose the network to non-evanescently decaying samples during training. This can be done using various criteria to select specific training samples from a large data pool. 
\textbf{(B)} A first selection is a totally random selection. In this case, the neural network is incapable of predicting lateral near-fields divergent from evanescence. This is a consequence of underexposure to these types of fields during training.
\textbf{(C)} A solution is to add more divergent samples to the training dataset. One way this can be accomplished is by selecting samples that do not have their maximum at the interface and adding these samples to the random data. This significantly improves the performance of the neural network for non-evanescent fields that do not have their maximum on the guide's interface (C center), but it still falls short for divergent fields that do have their maximum on the interface (C rightmost).
\textbf{(D)} Another approach is to exploit the difference in the spectrum between evanescent and non-evanescent fields. With this modification in place, the neural network can also predict the other types of divergent fields well (D rightmost).
}
\label{fig_4}
\end{figure}

We now proceed to train a neural network that predicts the lateral near-field landscape given a cascade of modes. This can be done completely analogously to the case for the longitudinal field. The only needed modification is a change of the output labels to $200$ points of the corresponding normalized lateral intensity profile. 
When training a network (Fig. \ref{fig_1}D) on a four-mode cascade one obtains a performance as shown in (Fig. \ref{fig_4}B). As can be seen, the network is very good at predicting classical evanescent fields (Fig. \ref{fig_4}B, leftmost). However, when a field starts to diverge from the typical monotonous decay the neural network starts to fail (Fig. \ref{fig_4}B, center and  rightmost). This is of particular importance since these types of lateral fields are the most interesting for some applications. The reason for this poor performance is that only a small fraction ($\approx 0.2\%$) of the randomly generated training data samples differ from the traditional evanescent behavior, making this type of behavior difficult to learn for the neural network. We found that a solution to this problem is to bias the network by artificially increasing the number of non-evanescent training samples. This can be done by selecting these non-evanescent samples from a much larger pool of random data samples and then adding these to the training dataset (Fig.~\ref{fig_4}A).
This data augmentation can be done in various ways. One method is to select all the near-fields that do not have their maximum on the waveguide's surface. These fields will not show simple monotonous decay away from the surface. When combining $25\%$ of these samples with $75\%$ random training samples, we can increase the exposure of the neural network to the divergent fields during training. Fig.~\ref{fig_4}C shows that this type of network is indeed better at predicting fields that diverge from evanescence and do not have their maximum on the guide's surface. Fields that diverge from the exponential decay but do have their maxima on the interface are still not well predicted (Fig. \ref{fig_4}C, rightmost example). Thus a better method is needed to distinguish all the non-evanescent samples from the evanescent ones. One way of tackling this is to exploit the difference in the spectrum of the non-evanescent profiles and the evanescent profiles. In other words, one can use the Fourier transform (FT) of the lateral profiles to augment the training data. Implementing this change and training a neural network on such an augmented dataset ($25\%$ selected, $75\%$ random) also improved performance for the other type of non-evanescent fields (Fig.~\ref{fig_4}D). Given the improved performance with this data augmentation step we will use such a neural network for the inverse design of lateral near-field landscapes. 
\section{The Inverse Design of Lateral Near-Field Landscapes}
\begin{figure}[htbp]
\centering
\includegraphics[width=\textwidth]{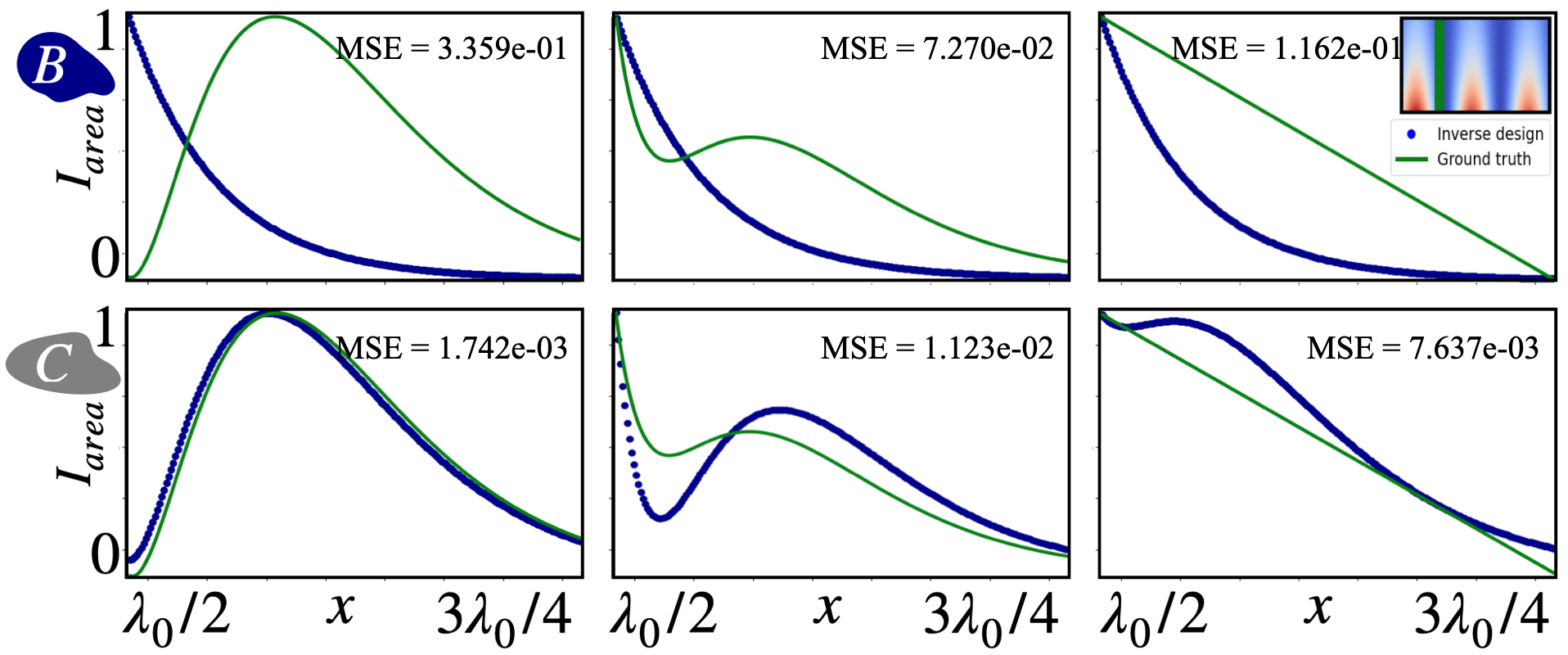} \\
\caption{\textbf{Inverse design of lateral fields} 
\textbf{(B)} The inverse design step for lateral field design, using a neural network trained on only random data (Fig. \ref{fig_4}B) results in poor performance. The main reason is an inadequately trained neural network. 
\textbf{(C)} When using a neural network trained on data partially selected with the Fourier criterion (Fig. \ref{fig_4}D), performance can be improved significantly. High sensitivity to the input parameters does remain, necessitating well-trained neural networks to obtain sufficient performance. 
}
\label{fig_5}
\end{figure}

The inverse design step here is analogous to the inverse design for the longitudinal fields. One uses a well-trained neural network (Fig. \ref{fig_1}D) and fixes the weights and biases of the hidden layers (Fig. \ref{fig_1}E). Next, one selects a desired field and runs the inverse design step, while limiting the parameter space.
Doing this using the network trained on only random data (Fig. \ref{fig_4}B) results in poor performance (Fig. \ref{fig_5}B), as can be expected since the neural network itself does not predict the optical response well for the non-evanescent samples (Fig. \ref{fig_4}B). However, using the network trained on the dataset partially selected using the Fourier transform criterion (Fig. \ref{fig_4}D) results in a significantly improved performance (Fig. \ref{fig_5}C). Note that the first two figures in (Fig. \ref{fig_5}C) are fields from the test dataset; it is thus known that these fields can be constructed using the cascade considered, and performance is therefore limited solely by the inverse design step. The rightmost field is an artificial field of which it is not known whether it can be constructed more optimally. 
Overall it was observed that the inverse design of lateral fields is harder than the case for longitudinal fields, given the need for the added data augmentation step during training as a consequence of the singular nature of the divergent fields in the randomly generated training dataset. Additionally it was observed that only a small difference in the design parameters could have a large effect on the optical response for non-evanescent samples, requiring a very well performing neural network, as a small discrepancy between the neural network's prediction and the actual landscape has a large effect on the inverse design step's performance. An example of this is shown in (Fig. \ref{fig_5}C, center); here the actual design parameters and those found by the inverse design step were very close, yet the small difference did result in a visible discrepancy. To further improve the inverse design step for fields like this, one would need to improve the neural network itself, by training longer on more data, or ideally by training smarter, to improve the sensitivity.  
By and large, the inverse design method, with the data augmentation step, did manage to construct many lateral fields successfully, proving to be a great improvement over other generic inverse design algorithms tested in ~\cite{ginis2020remote}.
\section{Conclusion}
To summarize, we successfully implemented the deep learning-based inverse design technique to the remote structuring of optical near-field landscapes. To realize this we imposed boundaries on the parameter space during the inverse design step so as to avoid nonphysical results. We also used an inversely-trained neural network to pose the initial guess. To design lateral fields we also artificially biased the training dataset, as to obtain networks capable of predicting non-evanescent fields. This last modification highlights the importance of a smart collection of training samples in order to obtain well-trained networks to run the inverse design step on. With these modifications in place, we managed to design for both longitudinal and lateral profiles of the near-field landscape.
Looking ahead, the scalability of the neural network-based inverse design approach opens up several exciting directions. The technique could be extended to create full 3D near-field landscapes by increasing the number of design parameters. One could extend the dimensionality of the beam even further and structure confined space-time beams~\cite{shiri2020hybrid, piccardo2022roadmap, ginis2020refracting, jolly2023coupling, yessenov2022space}. This would enable the construction of complex, physically relevant optical near-fields. Additionally, the data-driven nature of this framework makes it well-suited for designing nanophotonic components even when analytical solutions are unknown. This is particularly useful as many problems in nanophotonic design lack closed-form formulas. Further developing the inverse design methodology and integrating it with cascaded-mode optics could lead to a powerful platform for automated engineering of nanoscale optical fields and devices.
\section*{Acknowledgements}
JL acknowledges a fellowship from the Research Foundation Flanders (FWO - Vlaanderen) under Grant No. 11G1621N. VG acknowledges support from Research Foundation Flanders (FWO - Vlaanderen) under grant numbers G032822N and G0K9322N. 
\section*{Disclosures}
The authors declare no conflicts of interest.

\clearpage
\bibliographystyle{unsrtnat}
\bibliography{references}

\end{document}